\documentclass{emulateapj}
\usepackage{apjfonts}

\newcommand{\err}[2]{\ensuremath{^{_{+#1}}_{^{-#2}}}}
\newcommand{\ee}[2]{\ensuremath{{#1}\!\times\!10^{#2}}}
\newcommand{\hubble}{\textit{Hubble Space Telescope}}
\newcommand{\hst}{\textit{HST}}
\newcommand{\chandra}{\textit{Chandra}}

\newcommand{\xmm}{\textit{XMM}}
\newcommand{\xmmn}{\textit{XMM-Newton}}

\newcommand{\ergcms}{\ensuremath{\mathrm{erg\,cm^{-2}\,s^{-1}}}}
\newcommand{\Lx}{\ensuremath{L_\mathrm{x}}}
\newcommand{\Fx}{\ensuremath{F_\mathrm{x}}}
\newcommand{\kms}{\ensuremath{\mathrm{km\,s}^{-1}}}
\newcommand{\Msun}{\ensuremath{M_\sun}}
\newcommand{\ergsec}{\ensuremath{\mathrm{erg\,s}^{-1}}}
\newcommand{\pcmsq}{\ensuremath{\mathrm{cm}^{-2}}}

\begin{document}
%%%%%%%%%%%%%%%%%%%%%%%%%%%%%%%%%%%%%%%%%%%%%%
\shorttitle{X-rays from G1}
\shortauthors{Pooley \& Rappaport}
\slugcomment{accepted to ApJL}
%%%%%%%%%%%%%%%%%%%%%%%%%%%%%%%%%%%%%%%%%%%%%%
\title{X-rays from the Globular Cluster G1: Intermediate Mass Black Hole or Low Mass X-ray Binary?}

\author{David Pooley\altaffilmark{1}}
\affil{University of California at Berkeley\\ Astronomy Department, 601 Campbell Hall, Berkeley, CA 94720}
\email{dave@astron.berkeley.edu}

\and 
\author{Saul Rappaport}
\affil{Massachusetts Institute of Technology\\ Department of Physics and Kavli Institute for Astrophysics and Space Research, 70 Vassar St., Cambridge, MA 02139}
\email{sar@mit.edu}

\altaffiltext{1}{Chandra Fellow}

%%%%%%%%%%%%%%%%%%%%%%%%%%%%%%%%%%%%%%%%%%%%%%
\begin{abstract}
The globular cluster G1 (Mayall II) in M31 is the most massive ($\sim$$10^{7} M_\sun$) stellar cluster in the Local Group, and it has the highest central velocity dispersion ($\sim$28 \kms).  It has been claimed to host a central $\sim$20,000~\Msun\ black hole, but these claims have been controversial.  We report here the \xmmn\ detection of X-ray emission from G1 at a level of $\Lx\approx 2\times10^{36}$~\ergsec.  This emission could be the result of Bondi-Hoyle accretion of ionized cluster gas by a central black hole, or it could be produced by a conventional low-mass X-ray binary.  A precise localization of the X-ray emission, which is not possible with the current \xmm\ data, could distinguish between these possibilities.  While such a measurement may be difficult, it is of sufficient potential importance to pursue.

\end{abstract}

\keywords{globular clusters: individual (Mayall II = G1) --- X-rays: binaries}
%%%%%%%%%%%%%%%%%%%%%%%%%%%%%%%%%%%%%%%%%%%%%%

\section{Introduction}
\label{sec:intro}

Dynamical modeling of the kinematic data of numerous nearby galaxies has revealed that central dark objects of $10^6$--$10^{10}$ \Msun\ are nearly ubiquitous \citep[e.g.,][]{1995ARA&A..33..581K,1998AJ....115.2285M}.  These central dark objects are generally assumed to be supermassive black holes, the remnants of quasars which were much more numerous in the past.  Another population of black holes, those with masses of $\sim$3--20 \Msun, is known from determinations of the binary parameters of 18 X-ray binaries in the Milky Way and Magellanic Clouds \citep[e.g.,][]{2006RM}.  Bridging the gap between these two populations, i.e., finding intermediate mass black holes (IMBHs), has been a long-standing goal.

Some evidence for IMBHs has come from application of the dynamical modeling technique to globular clusters, which has found central dark objects in a couple of cases.  \citet{2002AJ....124.3270G} and \citet{2002AJ....124.3255V} claimed that the Milky Way globular cluster M15 hosted a central IMBH with a mass of $(3.9\pm2.2)\times10^{3}$~\Msun, and \citet{2002ApJ...578L..41G} argued that the Andromeda cluster G1 hosted an IMBH of $2\err{1.4}{0.8}\times10^4$~\Msun.  Both of these claims were challenged by \citet{2003ApJ...582L..21B,2003ApJ...589L..25B}.  

The case for M15 hosting a central IMBH was weakened by the discovery of a mislabeling of axes \citep{2003ApJ...585..598D}, but the case for G1 has recently been strengthened by new, high-quality data \citep{2005ApJ...634.1093G}.  Some relevant physical parameters for G1 are given in Table~\ref{tab:g1}.  \citet{2001AJ....122..830M} have suggested that G1 could be the stripped core of a dwarf elliptical galaxy.

If a globular cluster does indeed host a central IMBH, and if it contains some amount of ionized gas \citep[as has been observed in 47 Tuc;][]{2001ApJ...557L.105F}, one might reasonably expect the IMBH to accrete the gas via a Bondi-Hoyle-Lyttleton process \citep{1941MNRAS.101..227H,1944MNRAS.104..273B,1952MNRAS.112..195B} and become an observable X-ray and/or radio source \citep[e.g.,][]{1978ApJ...221..234G,2001ApJ...550..172P,2004MNRAS.351.1049M}.  

\citet{2003ApJ...587L..35H} reported that a search for a central X-ray source in a \chandra\ observation of M15 yielded a null result.  We report here on the detection of X-rays from G1 with \xmmn.  This X-ray detection may be emission from a central IMBH but is also consistent with emission from a low-mass X-ray binary (LMXB).  The observations are described in \S\,\ref{sec:obs} and the analysis in \S\,\ref{sec:analysis}.  We discuss the possible interpretations in \S\,\ref{sec:discuss} and briefly conclude in \S\,\ref{sec:conclusions}.

%%%%%%%%%%%%%%%%%%%%%%%%%%%%%%%%%%%%%%%%%%%%%%

\section{Observations}
\label{sec:obs}
\xmm\ has observed G1 on three occasions.  A log of the observations is given in Table~\ref{tab:obslog}.  We downloaded the reprocessed European Photon Imaging Camera (EPIC) data for each observation from the \xmm\ Science Archive\footnote{\url{http://xmm.vilspa.esa.es/external/xmm\_data\_acc/xsa}}.  EPIC consists of three separate cameras: one PN CCD array and two MOS CCD arrays.  The first observation had G1 at the aimpoint, but the other two included G1 at a considerable off-axis angle, which reduced the effective area due to vignetting.  To give a sense of this, we list in Table~\ref{tab:obslog} the effective area of the PN instrument at 1 keV, as determined from the ancillary response files.  The \xmm\ Optical Monitor was off for the first observation and did not include G1 in the field of view for the other two observations.

\begin{figure*}
\centering
\includegraphics[width=\textwidth]{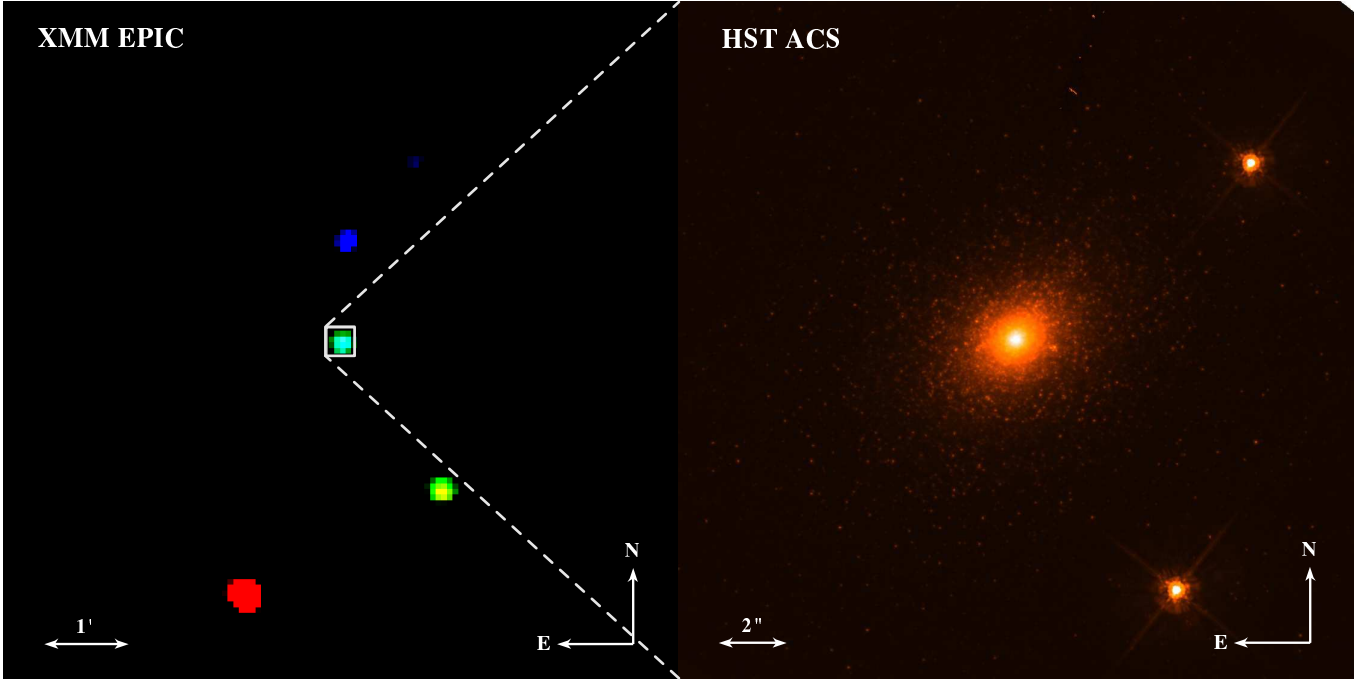}
\caption{{\it Left:} An $8'\times8'$ \xmmn\ image of G1 with photons in the range 0.5--1.2 keV shown in red, those in 1.2--2.5 keV shown in green, and those in 2.5--6 keV shown in blue.  The image has been smoothed by a Gaussian with a full width at half maximum of 12\arcsec\ and is displayed with a high contrast ratio in order to suppress the background. The small box indicates the area of the \hst\ image shown on the right. {\it Right:} A $20''\times20''$ \hst\ image of G1 taken with the F555W filter.}
\label{fig:xmmhst}
\end{figure*}

We used the \xmm\ Science Analysis System (SAS) version 6.5.0 (obtained from the \xmm\  Science Operations Centre\footnote{\url{http://xmm.esac.esa.int/}}) for the reduction of the data.  The second observation suffered from intense background flaring, and we filtered out the worst flaring periods to use only a few timeframes which were relatively unaffected.  This resulted in a significantly reduced exposure time.

We extracted spectra from the location of G1 for each instrument for each observation and constructed response matrices and ancillary response files using SAS.  We used large, nearby, source-free regions to extract background spectra.  The photometry based on these extractions is given in Table~\ref{tab:phot}.  We have combined the detected counts from the PN and both MOS instruments.  Because the exposure time and effective area vary greatly between instruments and observations, we report both net counts and net count flux.  We use the mean effective areas over 0.2 to 10 keV to calculate the fluxes.

An image of the first observation is shown in Figure~\ref{fig:xmmhst} in which the photons have been color-coded according to energy.  We also show a \hubble\ image of G1 taken with the Advanced Camera for Surveys with the F555W filter.  This multidrizzled image was downloaded from the Multimission Archive at the Space Telescope Science Institute\footnote{\url{http://archive.stsci.edu/}}.

%%%%%%%%%%%%%%%%%%%%%%%%%%%%%%%%%%%%%%%%%%%%%%

\section{Analysis}
\label{sec:analysis}

Since the count fluxes of all three observations are consistent with each other, we use only the first for spectral analysis because it has the best signal to noise ratio.  We jointly fit the three spectra (from the PN and both MOS instruments) from the first observation in Sherpa \citep{2001SPIE.4477...76F} using an absorbed power law model.  We considered only the 0.2--10 keV band, and the spectra were grouped to contain at least 3 counts per bin.  We used $\chi^2$ statistics with the \citet{1986ApJ...303..336G} estimate of the standard deviation of each bin.  The column density was constrained to be at least equal to the Galactic value of \ee{6.57}{20}~\pcmsq\ \citep{1990ARA&A..28..215D}.  The parameters of the best fit are given in Table~\ref{tab:spec}.  They are not well constrained.  The unabsorbed 0.2--10 keV luminosity is $\sim$\ee{2}{36}~\ergsec.

%%%%%%%%%%%%%%%%%%%%%%%%%%%%%%%%%%%%%%%%%%%%%%

\section{Discussion}
\label{sec:discuss}

\subsection{IMBH accreting cluster gas}
A black hole of mass $M_{\rm BH}$, moving with relative speed $v$ through a gas of ambient density $\rho$ and sound speed $c_s$, nominally accretes at the Bondi-Hoyle-Lyttleton rate (Hoyle \& Lyttleton 1941; Bondi \& Hoyle 1944; Bondi 1952; hereafter BHL) of:
\begin{equation}
\dot M \simeq 4\pi (GM_{\rm BH})^2\rho ({\mathsf v}^2+c_s^2)^{-3/2}~~~.
\end{equation}
If we define $V \equiv (v^2+c_s^2)^{1/2}$ and $\rho \equiv m_p n$, where $m_p$ is the proton mass and $n$ is the hydrogen number density, then for a black-hole mass of $10^4\,M_\odot$ the corresponding X-ray luminosity is given by:
\begin{eqnarray}
&\Lx \simeq \epsilon \dot Mc^2\simeq & \\
&5 \times 10^{37} \left(\frac{M_{\rm BH}}{10^4 M_\odot}\right)^2\left(\frac{V}{15\,{\rm km/s}}\right)^{-3}\left(\frac{n}{0.1}\right)~ {\rm ergs/s}&
\end{eqnarray}
where we have taken $\epsilon=0.057$ for the efficiency of converting mass to radiant energy in the inner disk of a Schwarzschild BH, and $V$ has been normalized to that expected for a gas at $T=10^4$ K with no significant bulk motion relative to the BH.  (Of course, the fraction of the energy emitted in the X-ray band may be considerably smaller than unity.) In order for hydrodynamic accretion to take place, the gas most likely has to be ionized (for mean free paths to be shorter than the accretion radius), and an ionized state may be achieved by the UV from only one or two post-AGB stars in the cluster.  \citet{2001ApJ...550..172P} argued that the wind from cluster AGB stars could maintain the intercluster gas density at $\sim$0.1--1 cm$^{-3}$, even after the cluster is completely stripped of its gas upon passing through the galactic plane of the host galaxy.  

Direct empirical evidence for a substantial ionized gas density in 47 Tuc, via observations of differential dispersion measures of the resident radio pulsars \citep{2001ApJ...557L.105F}, yields a value of $n = 0.067 \pm 0.015$ cm$^{-3}$ which confirms the possibility that globular clusters may indeed contain a non-negligible density of ionized gas.  %The implied value of \Lx\ with this density is quite comparable with the observed $\Lx \simeq 2 \times 10^{36}$ ergs s$^{-1}$ for the region containing G1.

The case of Bondi-Hoyle accretion by an IMBH is complicated by the large number of field stars that are likely to be contained within an accretion radius.  The accretion radius used to derive eq.\,(1) is:
\begin{equation}
R_{\rm acc} \simeq \frac{2GM_{\rm BH}}{V^2} \simeq 0.4 \left(\frac{M_{\rm BH}}{10^4 M_\odot}\right)\left(\frac{V}{15\,{\rm km/s}}\right)^{-2}~~{\rm pc}
\end{equation}
According to Fig.\,6 of \citet{2005ApJ...634.1093G}, there are more than $10^5$ stars within $\sim$0.4 pc of the center of G1.  Thus, we should consider both the dynamical effect of such stars on any gas present and the direct injection of gas into the accretion volume via the stellar winds of these stars.  We start by considering the collective effects of the stellar wind from the stars neighboring the IMBH.  Each star has a stellar wind intensity and speed that depend sensitively on its evolutionary state.  We assume that these winds collide and their bulk kinetic energies are thermalized.  

\begin{deluxetable}{lll}
\tablewidth{0pt}
\tablecaption{Properties of G1 in M31\label{tab:g1}}
\tablehead{
\colhead{Parameter} &\colhead{Value} &\colhead{Ref.}} 
\startdata
Total Mass ($M$) & $1.5 \times 10^7\,\Msun$ & [1]\\
Central Density ($\rho_0$) & $4.7\times10^5\,\Msun \,{\rm pc}^{-3}$ & [1]\\
Central Velocity Dispersion ($\sigma_0$) & 27.8 \kms & [1]\\
Core Radius ($r_c$) & $0.14''$ & [1]\\
Half-Mass Radius ($r_h$) & $3.7''$ & [1]\\
Tidal Radius ($r_t$) & $54''$ & [1]\\
Distance ($d$) & 750 kpc & [2]
\enddata
\tablerefs{[1] \citet{2001AJ....122..830M}; [2] \citet{2001ApJ...553...47F}}
\end{deluxetable}

In order to estimate the net mass loss and mean thermal velocities of the colliding winds, we consider the evolution of a single star near the turnoff mass (i.e., with $M \simeq 0.85\,M_\odot$).  We take the evolution of this one star to represent an ensemble of stars in different evolutionary states.  We utilized a simplified evolution code \citep{2000ApJ...538..241S} with a wind loss prescription that is a modified version of Reimers' (1975) expression.  The Reimers wind is enhanced as a function of the star's mass and radius (see eq.\,[15] in Soker \& Rappaport 2000) so as to reproduce the initial-final mass relation for the production of white dwarfs in single stars \citep{1993mlab.conf...55W}.  We take the terminal wind speed, $v_w$, to be equal to the escape speed from the stellar surface.  During the evolution of our representative cluster star we tabulated two integrals: 
\begin{equation}
\Delta KE = \int_{\rm ZAMS}^{\rm AGB} \frac{1}{2} \dot M v_w^2 dt~; ~~\Delta M = \int_{\rm ZAMS}^{\rm AGB} \dot M  dt~~.
\end{equation}
From these we compute the mass-weighted mean specific kinetic energy and the mean mass-loss rate from the winds:
\begin{eqnarray}
\frac{1}{2} <v_w^2> ~ & \simeq & ~\, \Delta KE/\Delta M \\
<\dot M> ~ & \simeq & ~\Delta M/\Delta t ~~~,
\end{eqnarray}
where $\Delta t$ is the evolution time from the ZAMS to the end of the AGB phase.

From these calculations we find the rms wind speed after collisional thermalization to be $v_{\rm w,rms} \equiv v_{\rm th} \simeq 55$ km s$^{-1}$, while the average value of $\dot M \simeq 2 \times 10^{-11}\,M_\odot$ yr$^{-1}$.  Of course, the bulk of the wind loss contribution comes from stars on the giant branch or AGB.  If we utilize this wind speed in eq.\,(4) we estimate $R_{\rm acc} \simeq 0.06$ pc.  Based on the central density $\rho_0$, there are $\sim$425 \Msun\ of stars in this volume, the bulk of which should be composed of the most massive cluster stars, near the turnoff or in the giant phase, which will have preferentially sunk to the cluster center.  These $\sim$500 stars could produce a total mass injection rate, via stellar winds, $\gtrsim$$10^{-8}\,M_\odot$ yr$^{-1}$.  This is approximately 100 times more than is needed to power the observed $L_x$ from the vicinity of G1.  Finally, we point out that since the rms speed of the cluster stars is lower than the rms thermal velocities that we estimate ($v_{\rm th} \simeq 55$ km s$^{-1}$), the dynamical effects of the stars on the gas are probably negligible.  

\subsection{LMXB(s) in outburst}

In the Milky Way, we know of 13 LMXBs in twelve globular clusters with $\Lx\ga10^{36}$~\ergsec\ \citep[for a review, see][and references therein]{2006VL}.  Five of these sources are recurrent transients, and the other seven are persistently luminous.  \citet{2003ApJ...591L.131P} have shown that the number of globular cluster close binaries observable as faint X-ray sources (e.g., LMXBs in quiescence, cataclysmic variables, active main-sequence binaries) scales as the ``encounter frequency'' of the cluster, which goes roughly as ${\rho_0}^{1.5} {r_c}^2$.  In their units, G1 has an encounter frequency of $\sim$7500 and would be expected to host $\sim$75 LMXBs, the vast majority in quiescence.  If some small number of those LMXBs were in outburst during each of the three \xmm\ observations, it could explain the observed X-ray emission.

\begin{deluxetable}{llllc}
\tablewidth{0pt}
\tablecaption{\xmmn\ Observations of G1 \label{tab:obslog}}
\tablehead{
\colhead{Date} & \colhead{ObsID} & \multicolumn{2}{c}{Exp.\ Time (s)} & \colhead{PN eff.\ area}\\
\colhead{}     & \colhead{}      & \colhead{PN}     & \colhead{MOS}   & \colhead{(cm$^2$ at 1 keV)} }
\startdata
2001 Jan 11& 0065770101& 4486.5 & 7325.1 & 894.2\\
2002 Dec 29& 0151580101& 1954.4\tablenotemark{\dag} & 3816.9\tablenotemark{\dag} & 482.1\\
2003 Feb 06& 0151581101& 7271.3 & 9563.6 & 556.4
\enddata
\tablenotetext{\dag}{Filtered for background flares.  The total exposure times were 6106.8 s for the PN and 8189.4 s for both MOS cameras.}
\end{deluxetable}

\subsection{Distinguishing the two possibilities}

Because the observed value of \Lx\ is compatible with either scenario, we discuss two possible ways to distinguish a central IMBH from an LMXB.  The first is by its location with respect to the cluster center, where the more massive the object the closer it is likely to be found to the center \citep[e.g.,][]{1979ApJ...231L.125J,1984ApJ...282L..13G}.  Near the cluster core, the stellar density approaches a constant value, and the cluster potential grows as $r^2$.  If black holes (BHs), neutron stars (NSs), and field stars are in rough energy equipartition, then we expect the rms black hole velocity to scale as $v_{\rm NS}\sqrt{M_{\rm NS}/M_{\rm BH}}$ where $M_{\rm NS}$ and $v_{\rm NS}$ are typical neutron star masses and velocities.  Thus, we would anticipate that any IMBH would be confined within a radius of $r_{\rm NS} \sqrt{M_{\rm NS}/M_{\rm IMBH}} \simeq 0.02\, r_{\rm NS}$, where $r_{\rm NS}$ is the radius containing about half the neutron stars in the cluster \citep[for a more detailed discussion see, e.g.,][]{1979ApJ...231L.125J,1976ApJ...209..214B}.  According to \citet{2003ApJ...598..501H}, the neutron star distribution in clusters follows that of the field stars, but with a characteristic radial scale that is $\sim$60\% that of the field stars.  Therefore, we expect half of the LMXBs in G1 to be within $\sim$$0.6\,r_h \simeq 2\farcs2$ of the center, but a 20,000 \Msun\ IMBH to be within 50 milliarcsec of the center.  Unfortunately, the spatial resolution of \xmm\ does not allow for such a test, but a \chandra\ observation could easily localize the X-ray emission to a fraction of an arcsecond. If there are at least a few X-ray/optical coincidences in the \chandra\ field of view, it should be possible to register the \chandra\ and \hst\ frames to 0\farcs1--0\farcs2 accuracy. With such a registration, an IMBH should be essentially coincident with the center of G1, and there is a good chance that any detected LMXB would be certifiably removed from the center.  The probabilities that a cluster LMXB would lie within 0\farcs3, 0\farcs2, and 0\farcs1 of the center are 0.16, 0.11, and 0.04, respectively.

\begin{deluxetable}{lllll}
\tablewidth{0pt}
\tablecaption{X-ray Photometry of G1 in 0.2--10 keV \label{tab:phot}}
\tablehead{
\colhead{Obs.} & \colhead{Net Counts} & \colhead{Net Flux (cnts cm$^{-2}$ s$^{-1}$)}}
\startdata
2001 Jan 11& $51.5\pm8.6 $&  $(1.02\pm0.17)\times10^{-5}$\\
2002 Dec 29& $21.9\pm14.4$&  $(1.51\pm0.99)\times10^{-5}$\\
2003 Feb 06& $20.0\pm12.8$&  $(0.48\pm0.31)\times10^{-5}$
\enddata
\end{deluxetable}

Although accretion disk spectra are notoriously difficult to calculate from first principles, an IMBH and an LMXB may have observably different spectra.  It has been suggested \citep[e.g.,][]{2003Sci...299..365K,2003ApJ...585L..37M} that a cool multicolor disk spectral component, indicating a relatively low inner disk temperature, might indicate the presence of an IMBH.  The sense of the effect is that the larger the mass of the BH accretor, the lower the temperature, $T_i$ of the inner edge of the disk, which scales as:
\begin{equation}
T_i \propto \left(M_{\rm BH}\dot M/r_i^3\right)^{1/4} \propto \dot M^{1/4}M_{\rm BH}^{-1/2}
\end{equation}
for a simple thin-disk model.  Since some Galactic BH sources, with $M_{\rm BH} \simeq 8\,M_\odot$ and $\dot M\simeq 10^{-8}\,M_\odot$ yr$^{-1}$ have $T_i \sim$ 1 keV, then an IMBH in G1 might have $T_i \sim$ 10 eV.  Thus, an association of the X-ray source with an IMBH, as opposed to an ordinary LMXB containing a neutron star or stellar-mass BH, could be made on the basis of a very soft observed spectral component.  We know of no reported observations of such a component in G1.

\begin{deluxetable}{ll}[t!]
\tablewidth{0pt}
\tablecaption{Absorbed Power Law Fit Parameters\label{tab:spec}}
\tablehead{
\colhead{Parameter} & \colhead{Value}}
\startdata
Column density ($n_H$) & \ee{6.6^{+82}}{20}~\pcmsq \\
Photon index ($\alpha$)& 1.2\err{1.4}{0.6} \\
Normalization at 1 keV& \ee{2.4\err{9.2}{1.3}}{-6} $\mathrm{cnts\,keV^{-1}\,cm^{-2}\,s^{-1}}$\\
$\chi^2$/d.o.f & 4.5/16 \\
Absorbed \Fx\tablenotemark{\dag}  & \ee{2.8}{-14}~\ergcms\\
Unabsorbed \Fx\tablenotemark{\dag}& \ee{3.0}{-14}~\ergcms\\
Unabsorbed \Lx\tablenotemark{\dag}& \ee{2.0}{36}~\ergsec
\enddata
\tablenotetext{\dag}{0.2--10 keV}
\end{deluxetable}

%%%%%%%%%%%%%%%%%%%%%%%%%%%%%%%%%%%%%%%%%%%%%%

\section{Conclusions}
\label{sec:conclusions}

We have reported on the detection of X-ray emission from the massive star cluster G1 in Andromeda, observed three times with \xmm.  There are two very different possibilities for the origin of the X-rays: an IMBH accreting cluster gas or one or more LMXBs in outburst.  Unfortunately, the \xmm\ data do not allow for a discrimination of these two scenarios.  Future observations may reveal that the source of the X-rays is displaced from the center of the cluster (indicating an LMXB) or found at its center (indicating an IMBH), and they may detect a very soft spectral component (supporting the presence of an IMBH). 

\acknowledgements 

DP gratefully acknowledges support provided by NASA through Chandra Postdoctoral Fellowship grant number PF4-50035 awarded by the Chandra X-ray Center, which is operated by the Smithsonian Astrophysical Observatory for NASA under contract NAS8-03060.  SR received some support from Chandra Grant TM5-6003X.

\end{document}